\documentclass[letterpaper]{article} 
\usepackage[submission]{aaai2026}  
\usepackage{times}  
\usepackage{helvet}  
\usepackage{courier}  
\usepackage[hyphens]{url}  
\usepackage{graphicx} 
\usepackage{natbib}  
\usepackage{caption} 
\usepackage{algorithm}
\usepackage{algorithmic}
\usepackage{booktabs}
\usepackage{amsmath}
\usepackage{multirow}
\usepackage{newfloat}
\usepackage{listings}
\DeclareCaptionStyle{ruled}{labelfont=normalfont,labelsep=colon,strut=off} 
\floatstyle{ruled}
\newfloat{listing}{tb}{lst}{}
\floatname{listing}{Listing}
\title{Multi-Agent Path Finding via Offline RL and LLM Collaboration}
\author{
    Merve Atasever,
    Matthew Hong,
    Mihir Nitin Kulkarni, \\
    Qingpei Li,
    Jyotirmoy V. Deshmukh \\
    \textnormal{Department of Computer Science, University of Southern California}
}
\affiliations{
    \textsuperscript{\rm 1}Association for the Advancement of Artificial Intelligence\\

    1101 Pennsylvania Ave, NW Suite 300\\
    Washington, DC 20004 USA\\
    proceedings-questions@aaai.org
}

\usepackage{bibentry}

\begin{document}

\maketitle
\begin{abstract}
Multi-Agent Path Finding (MAPF) poses a significant and challenging problem critical for applications in robotics and logistics, particularly due to its combinatorial complexity and the partial observability inherent in realistic environments. Decentralized reinforcement learning methods commonly encounter two substantial difficulties: first, they often yield self-centered behaviors among agents, resulting in frequent collisions, and second, their reliance on complex communication modules leads to prolonged training times, sometimes spanning weeks. To address these challenges, we propose an efficient decentralized planning framework based on the Decision Transformer (DT), uniquely leveraging offline reinforcement learning to substantially reduce training durations from weeks to mere hours. Crucially, our approach effectively handles long-horizon credit assignment and significantly improves performance in scenarios with sparse and delayed rewards. Furthermore, to overcome adaptability limitations inherent in standard RL methods under dynamic environmental changes, we integrate a large language model (GPT-4o) to dynamically guide agent policies. Extensive experiments in both static and dynamically changing environments demonstrate that our DT-based approach, augmented briefly by GPT-4o, significantly enhances adaptability and performance.

\end{abstract}

\section{Introduction}

Multi-Agent Path Finding (MAPF) addresses the fundamental challenge of guiding multiple autonomous agents to their respective goals without collisions, a problem widely recognized for its computational complexity and practical importance in logistics, warehousing, and robotic swarm management. While centralized approaches offer solutions using global planning algorithms (e.g., CBS, M*, ODrM*, LaCAM), they assume full observability of an environment and lack robustness to real-time dynamic changes \cite{cbs,Astar,Mstar,odrm, lacam}. This limitation severely restricts their applicability to realistic, partially observable environments where agents can only perceive their immediate surroundings.

In response, decentralized Multi-Agent Reinforcement Learning (MARL) methods have been developed to empower agents to independently navigate through local observations \cite{mamba, qmix, vdn, iql, qplex}. However, existing decentralized MARL approaches encounter significant difficulties: agents frequently develop self-centered behaviors leading to increased collisions, and the integration of complex inter-agent communication modules results in substantial computational overhead, which dramatically extends training periods to days or even weeks.

To overcome these persistent challenges, we propose a fundamentally different decentralized approach by leveraging an offline reinforcement learning setup. The Decision Transformer (DT) architecture inherently excels in modeling long-horizon dependencies, effectively addressing the credit assignment problem inherent in MAPF scenarios characterized by sparse and delayed rewards. By harnessing offline data, our DT-based agents drastically reduce training time requirements from weeks to a matter of hours without compromising performance.

Moreover, standard RL-based methods often falter when the environment undergoes dynamic alterations, which leads to agent indecision, oscillatory behaviors, or failure to adapt promptly. To address this critical issue, we pioneer the integration of a Large Language Model (LLM), specifically GPT-4o, to dynamically guide and adjust agent behaviors when encountering environmental changes. Unlike other communication-heavy methods, our integration of GPT-4o is concise and intermittent, providing short bursts of global awareness and strategic adaptability, which significantly improves agent performance in dynamic conditions \cite{gpt4}.

We rigorously validate our proposed framework across diverse experimental setups, including static scenarios and dynamically altered environments. In dynamic conditions, GPT-4o intervention ensures rapid real-time adjustments, directly navigating agents towards newly designated goals, thereby circumventing inefficient exploratory behaviors exhibited by purely DT-based agents. Our approach thus demonstrates a balanced and effective synergy between the fast, locally-informed policies of DT agents and the adaptive, globally-aware guidance provided by GPT-4o.

The contributions of our paper are:
\begin{itemize}
\item A novel decentralized offline reinforcement learning approach employing Decision Transformer to solve MAPF efficiently, significantly reducing training time from weeks to mere hours while maintaining robust performance.
\item Effective management of the credit assignment challenge in long-horizon MAPF tasks, specifically addressing scenarios with delayed positive rewards at the end.
\item The pioneering integration of GPT-4o for dynamic adaptation in MAPF, enabling significant performance improvements in environments subject to real-time changes.
\item Comprehensive experimental validation in both static and dynamic scenarios, clearly demonstrating the advantages and practicality of our DT+LLM approach for responsive and adaptive multi-agent systems.
\end{itemize}

\begin{figure*}[h]
\begin{center}
\includegraphics[width=0.65\linewidth]{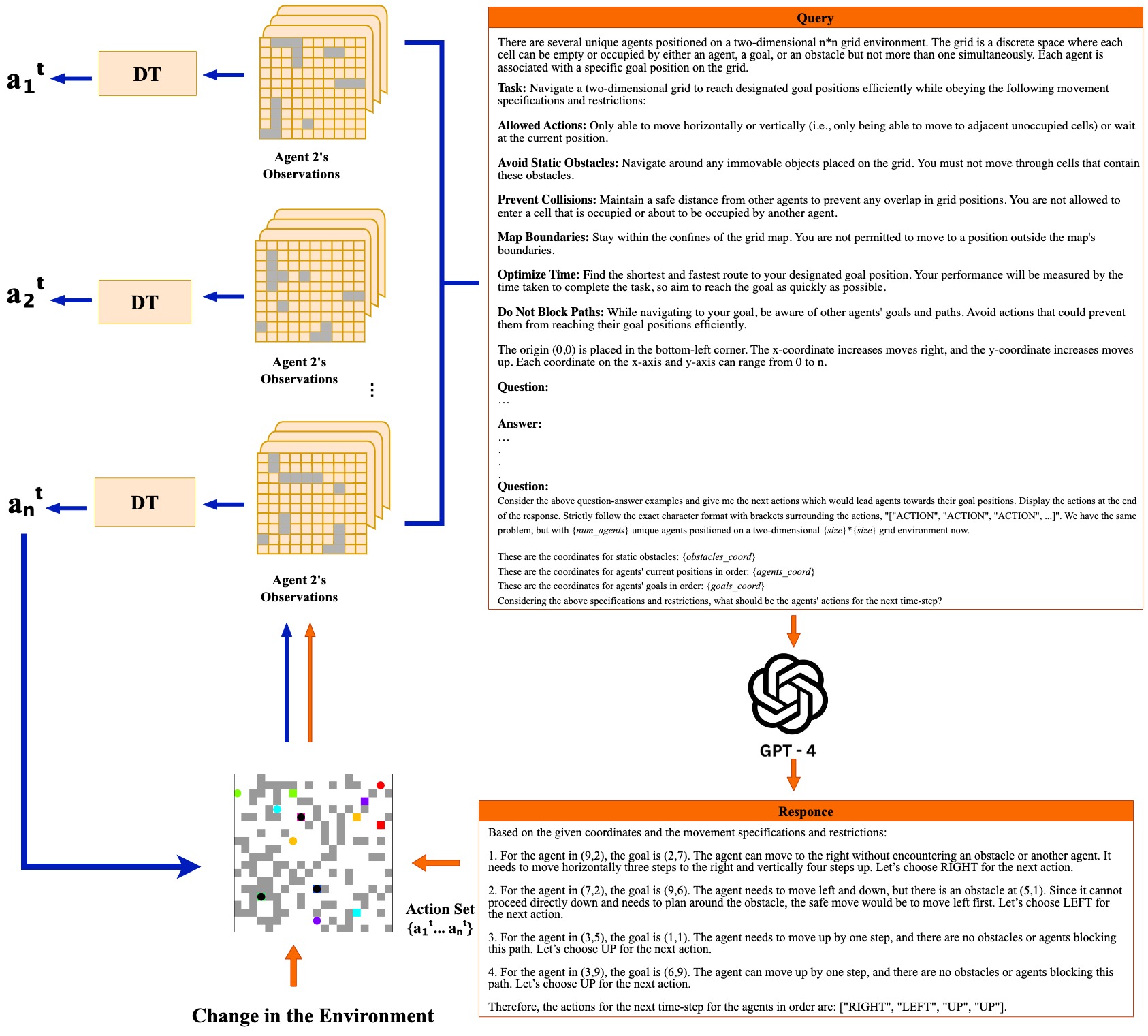}
\end{center}
\caption{Architecture of our pipeline: blue and orange arrows represent the processes of DT and GPT-4o, respectively.}
\label{pipeline}
\end{figure*}

\subsection{Related Work}

A prominent
benchmark is an approach called PRIMAL, where imitation learning (IL) is
combined with reinforcement learning (RL), allowing agents to {\em imitate}
centralized planner behaviors while being trained in a decentralized manner
\cite{primal}. Prioritized Communication Learning method (PICO) and PRIMAL$_2$ are extensions of
PRIMAL; whereas the first incorporates planning priorities and communication topologies into its
learning pipeline to improve collision avoidance and boost collaborative
behavior, the second works on lifelong version the problem \cite{pico, primal2}. On the other hand, Distributed Heuristic Learning with
Communication (DHC) tries to achieve this objective by employing graph
convolution for agent cooperation, \cite{dhc}. Here, each agent operates
independently with heuristic guidance provided through potential shortest-path
choices. Decision Causal Communication (DCC) enhances DHC by focusing on
selective communication. Unlike other methods, DCC enables agents to choose
relevant neighbors for communication, minimizing redundancy and overhead
\cite{dcc}. SCRIMP introduces a differentiable transformer-based communication
mechanism, which addresses the challenges posed by partial observations and
enhances team-level cooperation \cite{scrimp}. Finally, Confidence-based
Auto-Curriculum for Team Update Stability (CACTUS) proposes a reverse curriculum
approach that incrementally increases the potential distance between start and
goal locations to learn effective policy \cite{cactus}.

On the other hand, LLMs have demonstrated the ability to
understand and execute instructions for control and embodied tasks in robotics
\cite{etasks,robotics1,ecode,hu2023enabling,saycan}, coding
\cite{agentbench,evaluating}, strategic planning \cite{smartplay,llmp}, spatial
reasoning \cite{smartplay}, and even planning multi-agent collaboration tasks
\cite{theory,eagents,multic}. 
Interestingly, the direct
application of LLM-based agents does not yield comparable results to current
methods and underperforms the DT-based agents \cite{whynotsolving}.

\begin{figure*}[h]
\begin{center}
\includegraphics[width=0.7\linewidth]{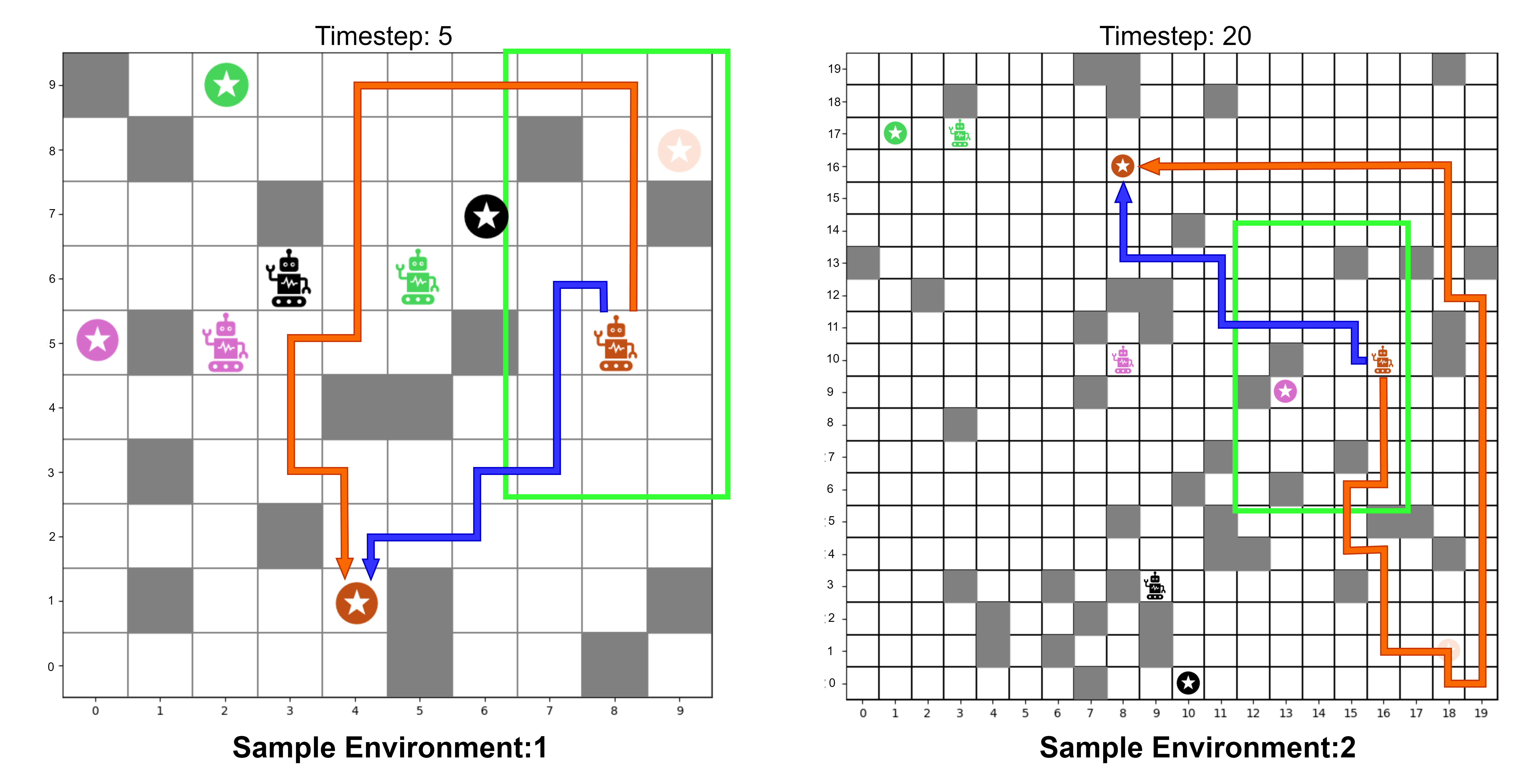}
\end{center}
\caption{Illustration of GPT-4o's assistance in the event of a goal change. In the first environment, the orange agent initially has its goal at (9,8), which is changed to (4,1) in the $5^{th}$ time step. In the second environment, the orange agent's goal is initially at (18,1), but is altered to (8,16) in the $20^{th}$ time step. The orange arrows depict the path generated by the DT alone, while the blue arrows represent the path taken when decision-making is switched to GPT-4o for five time steps after the goal change, before returning to DT. Green rectangles highlight the five time steps during which GPT-4o and DT make decisions. Without GPT-4o's assistance, DT agents initially explore the region around the previous goal before navigating to the new one. With GPT-4o's assistance, however, the agents can directly navigate toward the new goal.}
 \label{fig:goalchange}
\end{figure*}

Our experimental setup involves two scenarios: one with static environments and
one with altered environments. In the first case, DT agents are given a fixed
amount of time to finish the episode. GPT-4o is used if at least one of the DT
agents fails to complete the task within the allocated time. In the second
scenario, we change the goal positions of some agents during the inference and
include GPT-4o for only five timesteps after introducing the change in the
environment, then we switch the agent policy back to the DT-based policy. Here, we remark that integrating an LLM alters the partial observability setting to full observability (for five timesteps) in dynamically changing environments. This framework allows for modifications and real-time adjustments, enhancing the
adaptability of the agents as illustrated in Figure~\ref{fig:goalchange}.

\section{Preliminaries}

\subsection{Problem Setting}

Our research focuses on a deterministic and partially observable environment where a team of agents operates in a grid world to complete given tasks. Each agent is assigned to move from a start point to an endpoint and can either move to neighboring cells or remain stationary. The goal is achieved by all agents when they complete their tasks while minimizing the total time taken and avoiding collisions with obstacles or other agents. 

\subsection{Multi-Agent Path Finding}

Our setup constitutes a Decentralized Partially Observable Markov Decision Process (Dec-POMDP) which is a framework used in multi-agent systems where multiple agents must make decisions based on their individual observations of the environment \cite{decmarl}. The Dec-POMDP is defined by the tuple;
\[ (I, S, A, T, \Omega, O, R, \gamma ) \] consisting of \( I \): the set of agents, \( S \):  the set of states, \(A = \times_{i \in I} A_i\): the set of joint actions, where \(A_i\) is the set of actions available to agent \(i\), \(T(s' \mid s, a)\): the state transition function that gives the probability of transitioning to state \(s'\) from state \(s\) after joint action \(a\), \(\Omega = \times_{i \in I} \Omega_i\): the set of joint observations, where \(\Omega_i\) is the set of local observations of agent \(i\), \(O(o \mid s', a)\): the observation function that gives the probability of joint observation \(o\) given state \(s'\) and joint action \(a\), \(R(s, a)\): the reward function that gives the immediate reward received after taking joint action \(a\) in state \(s\), and \(\gamma\): the discount factor that determines the importance of future rewards.

In a shared environment, each agent behaves according to its policy \( \pi^i(a^i|s)=P(A^i_t=a^i| S_t=s) \), which is the probability distribution over actions given states. At a time step \( t \), the joint action \( \vec{a}(t) = (a^1_t, \ldots, a^N_t) \), where \( N = |I| \), leads to a transition to a new state \( s_{t+1}\) according to the transition function \( \ T \), and each agent receives a reward \( r^i_t \) according to the reward function \( R \). We consider a finite system where each episode plays out for a given \( \mathsf{T} \) timesteps. 

Additionally, we define \( G \) to be the set of goal states \( G \subseteq S \) and say that an agent $i$ has reached its goal if \(s^i_t \in G \) for some \(t \leq \mathsf{T} \). We consider an episode to terminate when all agents have reached their goals or at timestep \( \mathsf{T} \), whichever is sooner. 

\begin{table}[t]
\vspace{-0.1in}
\caption{Effort relative to other benchmarks in terms of training time, training episodes, and the number of parameters.}
\small
\label{effort}
\vspace{0.1in}
\begin{center}
\begin{tabular}{llll}
\multicolumn{1}{c}{\bf Benchmark}  &\multicolumn{1}{c}{\bf T. Time} &\multicolumn{1}{c}{\bf T. Episodes} &\multicolumn{1}{c}{\bf \# of Params}\\ 
\midrule
PRIMAL   & $\approx$ 20 days & 3.8M & 13M\\
DCC    & $\approx$ 1 day & 128K & 1M\\
DT     & $\approx$ 3 hours & 133K & 1.3M \\
\vspace{-0.3in}
\end{tabular}
\end{center}
\end{table}

\subsection{Decision Transformer}

The Decision Transformer treats offline RL as a sequence modeling problem and learns a return-conditioned policy from an offline dataset \cite{chen2021decision}. It has provided a novel perspective to reinforcement learning, and several extensions of this concept have been introduced subsequently \cite{onlinedt, multigamedt}. In the architecture, embedded tokens of returns, states, and actions are fed into a decoder-only transformer to generate the next tokens autoregressively using a causal self-attention mask. In other words, the model learns the probability of the next token \( x_t \) conditioned on previous \(K \) tokens, \( P_\theta(x_t|x_{t-K<... <t}) \), where \(K \)is a hyperparameter called \textit{context length}. To achieve this, we consider sequences of the form:

\[
\tau^i = (x_1^i, \ldots, x_t^i, \ldots, x_T^i) \quad \text{where} \quad x_t^i = (\hat{R}_t^i, o_t^i, a_t^i)
\]

such that \( \hat{R}_t^i \) is {\em return}-{\em to}-{\em go} (rtg) representing the cumulative rewards from the current time step until the end of the episode, \( o_t^i \) is the observation, and \(a_t^i \) is the action of agent \( i \)  at time \( t \). \\
We choose DT as the backbone for our method for three major reasons:
\begin{itemize}
    \item It is an offline RL algorithm that significantly reduces training time, as it does not require online interaction with the environment during training.
    \item The transformer architecture effectively addresses the credit assignment problem in long-horizon MAPF scenarios with positive rewards given only at the end. 
    \item DT performs well without extensive reward engineering by conditioning on the desired return.
\end{itemize}

\subsection{Large Language Models}

Building on the introduction of transformers in 2017, significant advancements in language models such as BERT, T5, the GPT series, Llama, and PaLM have extended the capabilities of LLMs and enabled their application to increasingly sophisticated tasks  \cite{attention, bert,t5,gpt1,gpt4,llama,palm}. Notably, these models function as few-shot learners for downstream tasks through prompt engineering without requiring further training \cite{surveyllm}. \\

\textbf{In-Context Learning:} (ICL) is an approach utilized by LLMs to handle downstream tasks by conditioning on relevant input-output examples or demonstrations \cite{icl, surveyicl}. Consequently, a pre-trained LLM model can tackle a wide range of tasks, from translation to question-answering, simply by modifying the examples in the prompt. This flexibility renders ICL a powerful tool for leveraging LLMs in new tasks. \\

\textbf{Chain of Thought:} (CoT) is an advanced demonstration design technique for prompt engineering used with LLMs to enhance their problem-solving abilities, particularly for tasks that require complex reasoning by structuring prompts to guide models through a step-by-step reasoning pathway \cite{cot}. Although several advancements in prompt engineering have emerged such as SayCan, ReAct, ToT, and other variations of CoT, we have chosen to utilize CoT in our work due to its simplicity and effectiveness \cite{saycan,react,tot,selfcot}.


\begin{figure*}[h]
\centering
\includegraphics[width=0.9\linewidth]{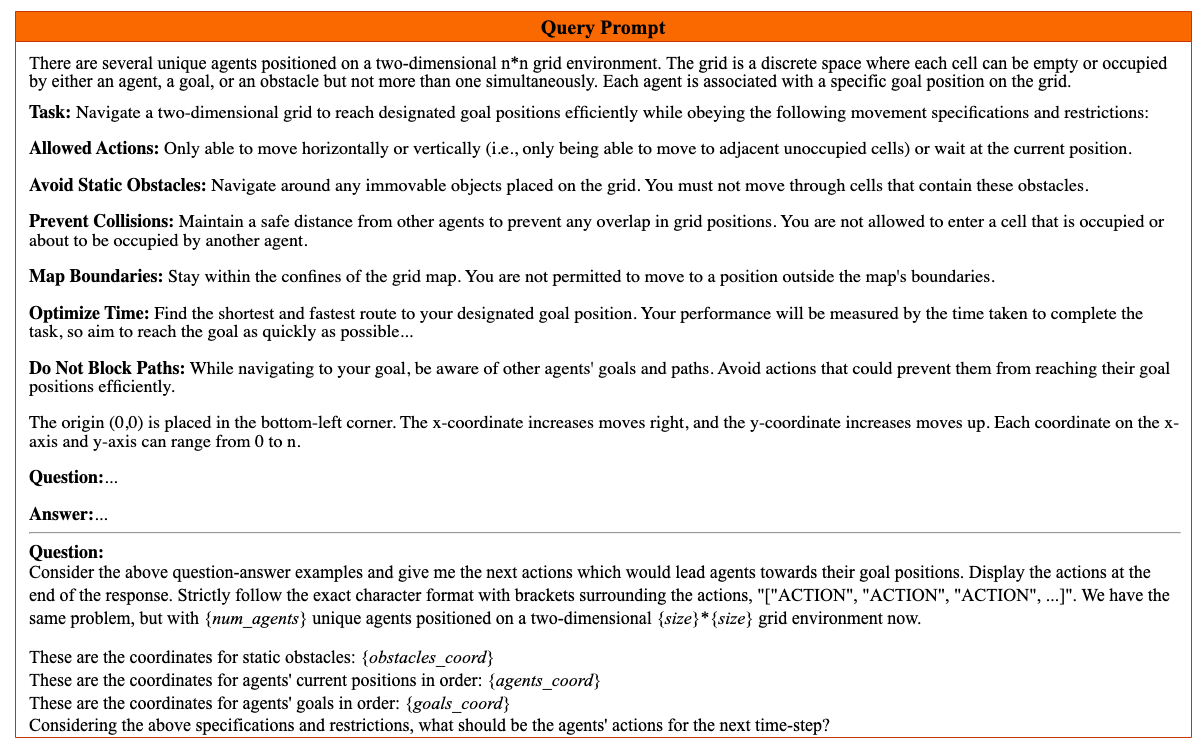}
\caption{In the query prompt, the MAPF problem, environment, and constraints are first outlined, followed by task-specific in-context examples provided using Chain-of-Thought (CoT).}
 \label{fig:prompt}
\end{figure*}

\section{Method}

We first create an offline dataset consisting of expert-level trajectories utilized to train the Decision Transformer. The trained DT model is deployed to each agent to generate the next action at each time step. The DT-based agents then navigate towards their goals within the test environment for predetermined time steps.
Subsequently, GPT-4o is integrated into the pipeline to assist agents that have yet to reach their goals. Environmental information, including the coordinates of static obstacles, the agents' current positions, and their target destinations, is encoded into a query-prompt. The prompt is then provided to GPT-4o which generates the next set of actions for one time-step. This iterative process continues for five time steps, after which the pipeline reverts to the DT policies to complete the episode. Detailed descriptions of the methodology and experimental results are presented in the following sections. The overall architecture of our pipeline is depicted in Figure \ref{pipeline}.

\subsection{Building Expert Trajectories \& Training}

To generate \textbf{expert-level} behavior, we collected trajectories using the ODrM* algorithm, a centralized classical MAPF solver frequently employed in the literature to create expert trajectories for imitation learning. The algorithm was executed on \( 80 \) randomly generated grid environments for each combination of varying parameters:  \( (4, 16, 32, 64) \) agents, grid sizes of \( (10, 20, 40, 80) \), and obstacle densities of  \( (0, 0.1, 0.2) \).

Each agent's path constitutes a distinct trajectory in the training dataset. Given the paper's emphasis on partially observable environments, agents are constrained to observing only their own fields of view (FOV), each of size \( 10 \times 10 \). Observations are represented by four 2-dimensional arrays of shape \( (10,10) \), encoding the following information about their local environments: positions of neighboring agents within the agent’s FOV, position of the agent’s own goal, positions of neighbors’ goals within the agent’s FOV, positions of obstacles within the agent’s FOV. Here, if the agent's goal lies outside its field of view, the goal is projected onto the edge cell closest to it. If the goal falls within the agent’s FOV, it is displayed in the corresponding cell.

At each time step, agents have the option to either wait or move in one of four directions (N/E/S/W) while receiving rewards as outlined:\(-0.3\) for moving, \((0/-0.5)\) for waiting (on/off goal), \(-5\) for collision, and \(+20\) for reaching the goal. We also created a modified version of the dataset in which agents receive an extra reward of \(+20\) upon successfully completing an episode, (i.e. all agents reach respective goals). However, the DT model trained on this modified dataset demonstrated inferior performance compared to the model trained on the original version.

Once the trajectories are collected, the dataset undergoes pre-processing to align with the input format required by the DT model. With the context length for the DT set to 50, we divided the trajectories into chunks of length 50. For chunks shorter than the specified context length, we applied zero-padding. The final dataset, composed of these chunks, was derived from a total of \(133\)K episodes.
\\
For the most part, we retained the original architecture of the Decision Transformer except for replacing the linear layer with a convolutional encoder to process observations of shape \( (4, 10, 10) \). 

\subsection{Prompting GPT-4}

We started with an initial prompt and iteratively refined it by incorporating corrective feedback from GPT-4o itself. While we experimented with prompt generation tools like AdalFlow\footnote{https://github.com/SylphAI-Inc/AdalFlow}, we found that our custom-designed prompts performed better overall. The prompt design that yielded the best performance has been integrated into our pipeline; the prompt template is illustrated in Figure \ref{fig:prompt}. \\
In the query prompt, we begin by outlining the problem, environment specifications, constraints, and the task. Following this, we provide task-specific in-context examples and pose a similar question generated by our pipeline. To construct these in-context examples, we analyzed environments where the DT model failed to find a path to a goal for at least one agent within \( \mathsf{T} \) timesteps. Using both a simple and a challenging (failed) scenario, we curated sample question-answer pairs.

\section{Experimental Results}

Model training and experiments were performed on an NVIDIA Quadro RTX \(5000\) with \(16\) GB GPU memory.

\subsection{Stationary Environments}

\begin{figure}[h]
\centering
\includegraphics[width=0.8\linewidth]{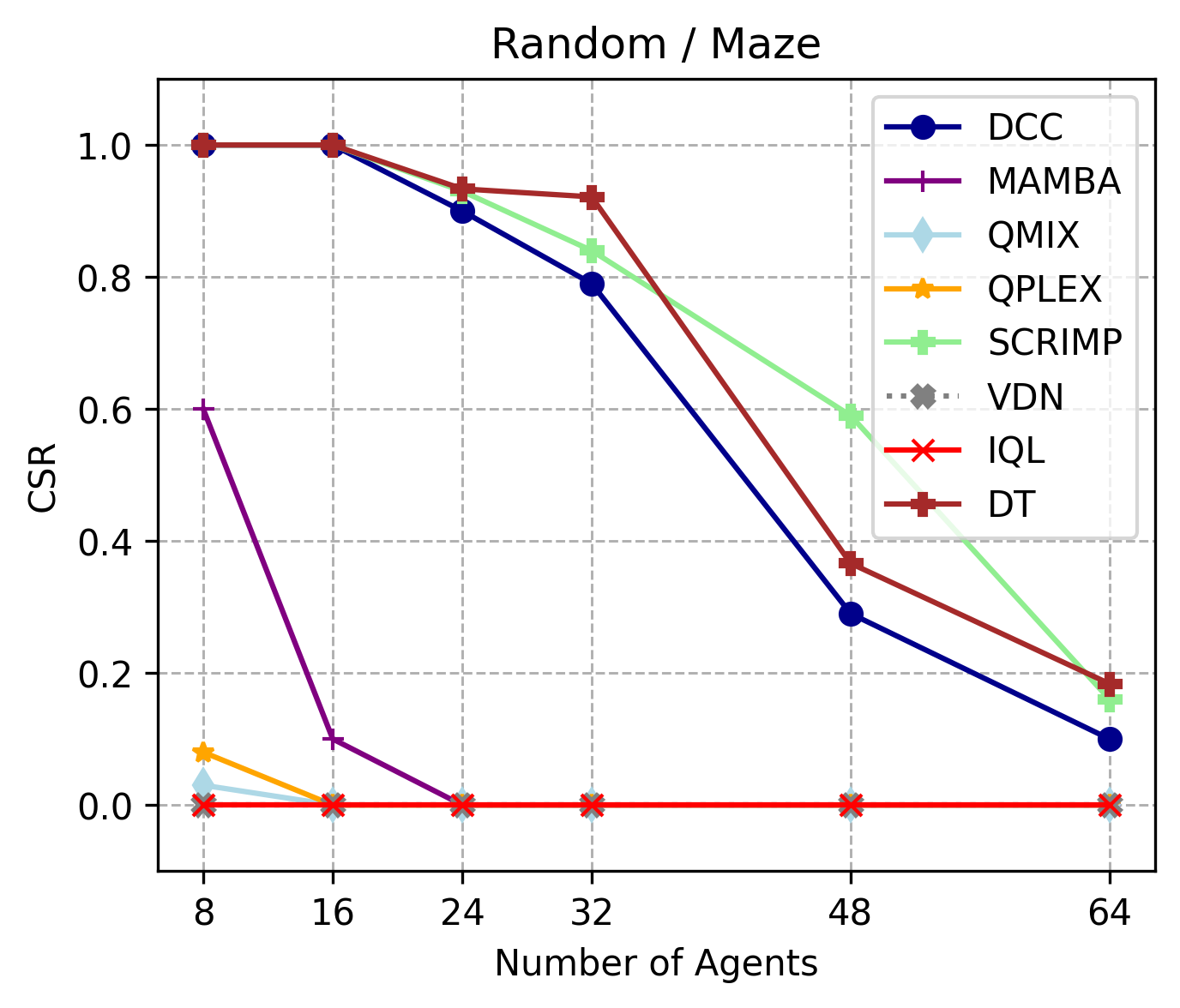}
\caption{Performance of MAPF methods on Random and Mazes maps; higher CSR is
better.}
 \label{fig:pogema1}
\end{figure}

\begin{figure}[h]
\centering
\includegraphics[width=0.8\linewidth]{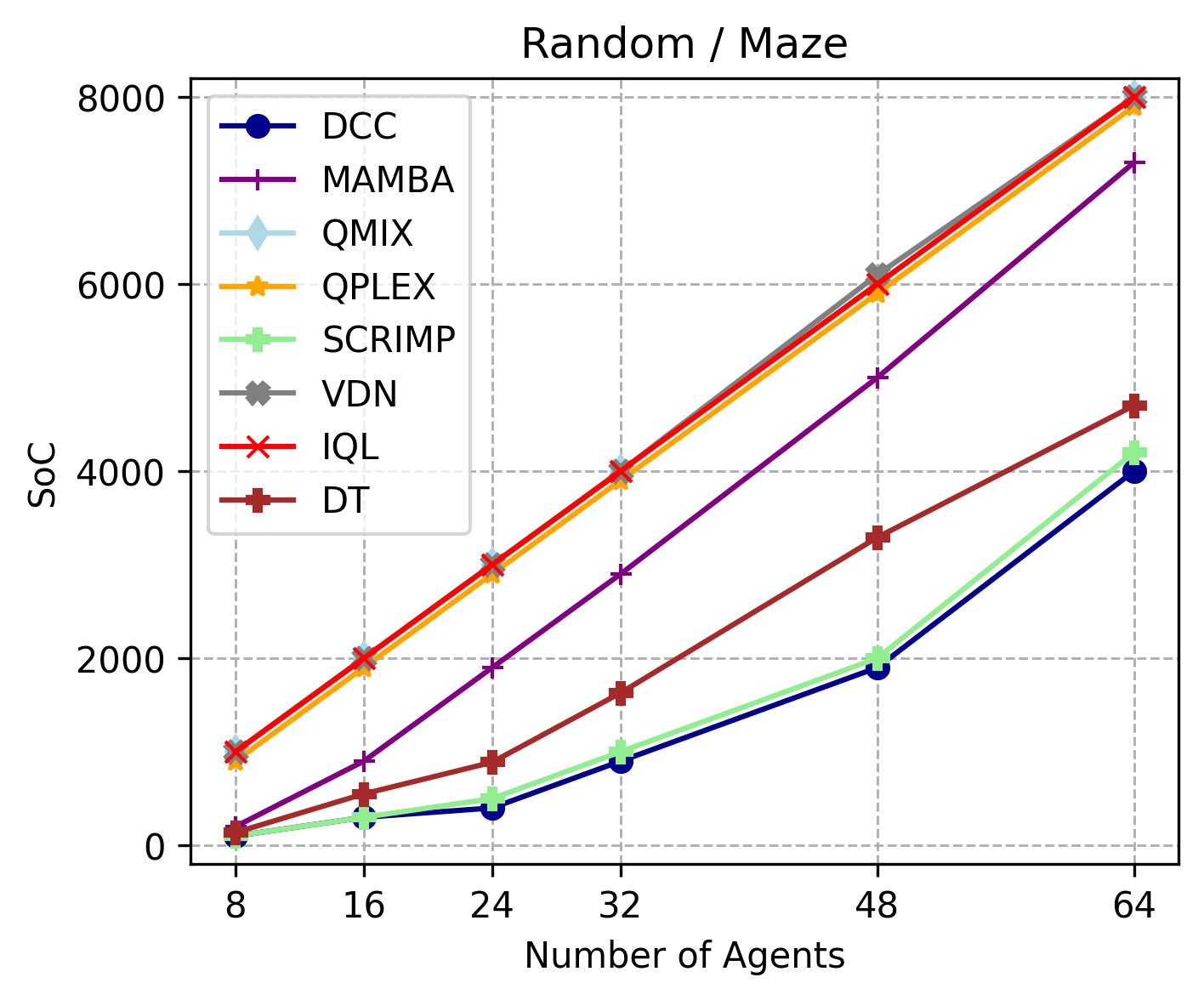}
\caption{Performance of MAPF methods on Random and Mazes maps; lower SoC is
better.}
 \label{fig:pogema2}
\end{figure}

The first part of the experiments are carried out by using \textit{Random} and \textit{Mazes} maps of the unified framework, Pogema, to be able to compare our method with other learning-based MAPF benchmarks \cite{pogema, pogemabench}. The results in Figure-\ref{fig:pogema1} and Figure-\ref{fig:pogema2} are obtained by running tests on the maps consist of 
\(\{ 17, 21\}\)-size grid environments with varying numbers of agents \( \{ 8, 16, 24, 32, 48, 64 \} \).
 The performance of the benchmarks is evaluated using two metrics: sum-of-costs (SoC, i.e. the sum of time steps across all agents to reach their respective goals) and cooperative success rate (CSR).

We draw the following conclusions from our experiments:
\begin{enumerate}
\item DT-based agents consistently outperform all baseline methods with higher success rates across all settings, except for the case with 48 agents. Notably, increasing the agent count to 64 yields the best overall performance.
\item SCRIMP demonstrates strong performance in Pogema's environments; however, the original paper does not report results for larger environments. The published findings are limited to environment sizes of 10, 30, and 40, indicating that SCRIMP may be optimized primarily for smaller-scale settings.
\item DT agents tend to take slightly longer paths compared to DCC and SCRIMP, yet they achieve a lower sum of costs, reflecting more globally efficient navigation behaviors.
\end{enumerate}

\subsection{{Dynamic Scenario Adaptation}}
The second part of our experiments are carried out in randomly generated \( \{20, 40, 80 \} \)-sized grid environments with varying numbers of agents \( \{ 8, 16, 32, 64 \} \), and obstacle densities \( \{ 0, 0.1, 0.2 \} \). The result, summarized in the Table \ref{Dynamic_Environments}, represent averages across 3 different obstacle density values and 60 environments (combinations of grid size, number of agents), resulting in a total of 720 test environments. We evaluated the performance of the benchmarks and our models using three key metrics: success rate (CSR or SR), makespan (MS), and collision rate (CR). The makespan refers to the duration of an episode, specifically the time taken for the last agent to reach its designated goal. Finally, the collision rate is computed as the number of collisions among agents in a successful episode divided by the episode's duration. Collisions occurring in unsuccessful episodes are excluded from this calculation.

\begin{table*}[t]
\vspace{-0.1in}
\caption{Impact of GPT-4o on dynamic environments: For \( 20, 40, 80 \) size environments, we modify the environment once at \( 15 \)th, \( 30 \)th, and \( 50 \)th timesteps respectively and conduct our experiments according to two difficulty levels; altering the goals of \( .25 \)  of the agents  and \(  .5 \) of the agents in the environments during inference. For makespan and collision rate, lower values (indicated by arrows) means better performance.}
\label{Dynamic_Environments}
\small
\centering
\vspace{0.1in}
\resizebox{\linewidth}{!}{%
\begin{tabular}{lc ccc ccc ccc ccc} 
 & & \multicolumn{3}{c}{\textbf{DT (1/4)}} & \multicolumn{3}{c}{\textbf{DT+GPT-4o (1/4)}} & \multicolumn{3}{c}{\textbf{DT (1/2)}} & \multicolumn{3}{c}{\textbf{DT+GPT-4o (1/2)}} \\
\cmidrule(lr){3-5} \cmidrule(lr){6-8} \cmidrule(lr){9-11} \cmidrule(lr){12-14}
\textit{Env Size} & \textit{\# Agents} & \textit{MS} \( \downarrow \) & \textit{SR}(\%) & \textit{CR} \( \downarrow \) & \textit{MS} \( \downarrow \) & \textit{SR} (\%) & \textit{CR}\( \downarrow \) & \textit{MS}\( \downarrow \) & \textit{SR} (\%) & \textit{CR}\( \downarrow \) & \textit{MS}\( \downarrow \) & \textit{SR} (\%) & \textit{CR}\( \downarrow \) \\
\midrule
\multirow{3}{*}{20x20} & 8  & 48.5  & 95 & 0.21 & 44.7  & 97 & 0.11 & 48.9  & 96 & 0.14 & 52.8  & 97 & 0.08\\
                       & 16 & 68.5  & 85 & 0.48 & 68.1  & 88 & 0.51 & 71.5  & 85 & 0.49 & 74.2  & 95 & 0.37\\
                       & 32 & 100.1 & 73 & 1.68 & 101.1 & 71 & 1.64 & 112.1 & 78 & 1.68 & 102.2 & 75 & 1.55\\
\midrule
\multirow{4}{*}{40x40} & 8  & 80.9  & 98 & 0.02 & 77.0  & 100 & 0.02 & 92.4  & 96 & 0.04 & 91.8  & 100 & 0.03 \\
                       & 16 & 97.2  & 98 & 0.15 & 92.4  & 98 & 0.14 & 104.4 & 92 & 0.13 & 97.1  & 98 & 0.12\\
                       & 32 & 126.4 & 82 & 0.67 & 125.5 & 82 & 0.50 & 120.2 & 75 & 0.48 & 120.2 & 82 & 0.54\\
                       & 64 & 162.9 & 60 & 2.03 & 150.7 & 61 & 1.82 & 172.7 & 60 & 2.18 & 161.5 & 64 & 1.82\\
\midrule
\multirow{4}{*}{80x80} & 8  & 133.0 & 94 & 0.01 & 128.5 & 94 & 0.00 & 147.9 & 93 & 0.02 & 145.4 & 92 & 0.01\\
                       & 16 & 144.4 & 92 & 0.03 & 141.0 & 97 & 0.03 & 146.0 & 90 & 0.04 & 150.1 & 94 & 0.07\\
                       & 32 & 155.5 & 82 & 0.18 & 158.0 & 81 & 0.17 & 171.7 & 78 & 0.16 & 166.3 & 85 & 0.24\\
                       & 64 & 182.4 & 66 & 0.59 & 176.3 & 62 & 0.72 & 169.8 & 63 & 0.70 & 183.0 & 62 & 0.53\\
\bottomrule
\end{tabular}}
\end{table*}

Based on the results presented in Table \ref{Dynamic_Environments}, we can draw the following conclusions:

\begin{enumerate}
\item 
The integration of GPT-4o reduces makespan across most environmental settings.
This reduction is particularly significant when the new goal location is in the
opposite direction to the agents' prior trajectory. As illustrated in Figure-\ref{fig:goalchange}, DT requires several timesteps to comprehend the goal change,
often exploring areas near the previous goal location before adjusting its
direction. In contrast, when the LLM-based suggestions are introduced
concurrently with the dynamic goal change, agents immediately reorient towards
their new goal location. This significantly reduces the makespan.
\item 
The success rates achieved by DT and LLM collaboration are equal to or surpass those of the DT alone in most environmental settings. Notably, when we alter the goal positions of half of the agents, the advantages of LLM guidance become particularly evident in complex environments characterized by larger sizes and a greater number of agents. 
\item
 DT \& GPT-4 collaboration surpasses DT in terms of collision rates, indicating that LLM integration results in safer behavior. This is particularly significant since collisions among agents in real-world scenarios, such as warehouses, can lead to substantial costs and safety hazards.

\end{enumerate} 

These findings demonstrate that LLM-assisted DT-based agents are highly
effective for real-time adaptations and offer significant advantages in
safety-critical environments.

\subsection{Discussion}


\textbf{Offline vs. Online RL Approaches in MAPF} We show that incorporating offline RL in MAPF (through the DT architecture) is an effective learning strategy that yields performance comparable to other learning-based methods that require online interaction with the environment during training. Table \ref{effort} highlights the significant reduction in required effort and our success in eliminating the necessity for real-time interaction with environments during training. Notably, our model does not experience the distributional shift issues that challenge offline RL algorithms when tested in new environments. By utilizing a dataset consisting of a broad range of samples from randomly generated grid environments, we mitigate the risk of distributional shift (analogous to \cite{domainrand}).



\section{Conclusion}

Our flexible framework introduces offline RL to MAPF into the literature and accommodates both static and dynamic environments, rapidly adapting to changes such as shifting goals. Although LLMs can occasionally hallucinate, i.e., yield outputs that deviate from factual accuracy or contextual relevance, our work seeks to harness the capabilities of LLMs within MAPF and points out contexts wherein the utilization of the models addresses specific challenges \cite{kambhampati}.


\textbf{Limitations \& Future Work} In this paper, we opted for textual inputs because LLMs are still largely unexplored within the MAPF literature. Replicating this approach using visual inputs and Visual Language Models (VLMs) presents a promising direction for future research. 
Given the rapid developments, we believe that the integration of LLMs into MAPF methods is promising.

\bibliography{aaai2026}

\begin{thebibliography}{52}
\providecommand{\natexlab}[1]{#1}

\bibitem[{Achiam et~al.(2023)Achiam, Adler, Agarwal, Ahmad, Akkaya, Aleman, Almeida, Altenschmidt, Altman, Anadkat et~al.}]{gpt4}
Achiam, J.; Adler, S.; Agarwal, S.; Ahmad, L.; Akkaya, I.; Aleman, F.~L.; Almeida, D.; Altenschmidt, J.; Altman, S.; Anadkat, S.; et~al. 2023.
\newblock Gpt-4 technical report.
\newblock \emph{arXiv preprint arXiv:2303.08774}.

\bibitem[{Ahn et~al.(2022)Ahn, Brohan, Brown, Chebotar, Cortes, David, Finn, Fu, Gopalakrishnan, Hausman et~al.}]{saycan}
Ahn, M.; Brohan, A.; Brown, N.; Chebotar, Y.; Cortes, O.; David, B.; Finn, C.; Fu, C.; Gopalakrishnan, K.; Hausman, K.; et~al. 2022.
\newblock Do as i can, not as i say: Grounding language in robotic affordances.
\newblock \emph{arXiv preprint arXiv:2204.01691}.

\bibitem[{Brown et~al.(2020)Brown, Mann, Ryder, Subbiah, Kaplan, Dhariwal, Neelakantan, Shyam, Sastry, Askell et~al.}]{icl}
Brown, T.; Mann, B.; Ryder, N.; Subbiah, M.; Kaplan, J.~D.; Dhariwal, P.; Neelakantan, A.; Shyam, P.; Sastry, G.; Askell, A.; et~al. 2020.
\newblock Language models are few-shot learners.
\newblock \emph{Advances in neural information processing systems}, 33: 1877--1901.

\bibitem[{Chang et~al.(2024)Chang, Wang, Wang, Wu, Yang, Zhu, Chen, Yi, Wang, Wang et~al.}]{surveyllm}
Chang, Y.; Wang, X.; Wang, J.; Wu, Y.; Yang, L.; Zhu, K.; Chen, H.; Yi, X.; Wang, C.; Wang, Y.; et~al. 2024.
\newblock A survey on evaluation of large language models.
\newblock \emph{ACM Transactions on Intelligent Systems and Technology}, 15(3): 1--45.

\bibitem[{Chen et~al.(2021{\natexlab{a}})Chen, Lu, Rajeswaran, Lee, Grover, Laskin, Abbeel, Srinivas, and Mordatch}]{chen2021decision}
Chen, L.; Lu, K.; Rajeswaran, A.; Lee, K.; Grover, A.; Laskin, M.; Abbeel, P.; Srinivas, A.; and Mordatch, I. 2021{\natexlab{a}}.
\newblock Decision transformer: Reinforcement learning via sequence modeling.
\newblock \emph{Advances in neural information processing systems}, 34: 15084--15097.

\bibitem[{Chen et~al.(2021{\natexlab{b}})Chen, Tworek, Jun, Yuan, Pinto, Kaplan, Edwards, Burda, Joseph, Brockman et~al.}]{evaluating}
Chen, M.; Tworek, J.; Jun, H.; Yuan, Q.; Pinto, H. P. d.~O.; Kaplan, J.; Edwards, H.; Burda, Y.; Joseph, N.; Brockman, G.; et~al. 2021{\natexlab{b}}.
\newblock Evaluating large language models trained on code.
\newblock \emph{arXiv preprint arXiv:2107.03374}.

\bibitem[{Chen, Koenig, and Dilkina(2024)}]{whynotsolving}
Chen, W.; Koenig, S.; and Dilkina, B. 2024.
\newblock Why Solving Multi-agent Path Finding with Large Language Model has not Succeeded Yet.

\bibitem[{Chowdhery et~al.(2023)Chowdhery, Narang, Devlin, Bosma, Mishra, Roberts, Barham, Chung, Sutton, Gehrmann et~al.}]{palm}
Chowdhery, A.; Narang, S.; Devlin, J.; Bosma, M.; Mishra, G.; Roberts, A.; Barham, P.; Chung, H.~W.; Sutton, C.; Gehrmann, S.; et~al. 2023.
\newblock Palm: Scaling language modeling with pathways.
\newblock \emph{Journal of Machine Learning Research}, 24(240): 1--113.

\bibitem[{Damani et~al.(2021)Damani, Luo, Wenzel, and Sartoretti}]{primal2}
Damani, M.; Luo, Z.; Wenzel, E.; and Sartoretti, G. 2021.
\newblock PRIMAL $ \_2 $: Pathfinding via reinforcement and imitation multi-agent learning-lifelong.
\newblock \emph{IEEE Robotics and Automation Letters}, 6(2): 2666--2673.

\bibitem[{Devlin et~al.(2018)Devlin, Chang, Lee, and Toutanova}]{bert}
Devlin, J.; Chang, M.-W.; Lee, K.; and Toutanova, K. 2018.
\newblock Bert: Pre-training of deep bidirectional transformers for language understanding.
\newblock \emph{arXiv preprint arXiv:1810.04805}.

\bibitem[{Dong et~al.(2022)Dong, Li, Dai, Zheng, Wu, Chang, Sun, Xu, and Sui}]{surveyicl}
Dong, Q.; Li, L.; Dai, D.; Zheng, C.; Wu, Z.; Chang, B.; Sun, X.; Xu, J.; and Sui, Z. 2022.
\newblock A survey on in-context learning.
\newblock \emph{arXiv preprint arXiv:2301.00234}.

\bibitem[{Egorov and Shpilman(2022)}]{mamba}
Egorov, V.; and Shpilman, A. 2022.
\newblock Scalable multi-agent model-based reinforcement learning.
\newblock \emph{arXiv preprint arXiv:2205.15023}.

\bibitem[{Ferner, Wagner, and Choset(2013)}]{odrm}
Ferner, C.; Wagner, G.; and Choset, H. 2013.
\newblock ODrM* optimal multirobot path planning in low dimensional search spaces.
\newblock In \emph{2013 IEEE International Conference on Robotics and Automation}, 3854--3859. Karlsruhe, Germany.

\bibitem[{Hart, Nilsson, and Raphael(1968)}]{Astar}
Hart, P.~E.; Nilsson, N.~J.; and Raphael, B. 1968.
\newblock A formal basis for the heuristic determination of minimum cost paths.
\newblock \emph{IEEE transactions on Systems Science and Cybernetics}, 4(2): 100--107.

\bibitem[{Hu et~al.(2023)Hu, Zhao, Zhang, Zhou, Yang, Xu, and Liu}]{hu2023enabling}
Hu, B.; Zhao, C.; Zhang, P.; Zhou, Z.; Yang, Y.; Xu, Z.; and Liu, B. 2023.
\newblock Enabling intelligent interactions between an agent and an LLM: A reinforcement learning approach.
\newblock \emph{arXiv preprint arXiv:2306.03604}.

\bibitem[{Kambhampati(2024)}]{kambhampati}
Kambhampati, S. 2024.
\newblock Can large language models reason and plan?
\newblock \emph{Annals of the New York Academy of Sciences}, 1534(1): 15--18.

\bibitem[{Lee et~al.(2022)Lee, Nachum, Yang, Lee, Freeman, Guadarrama, Fischer, Xu, Jang, Michalewski et~al.}]{multigamedt}
Lee, K.-H.; Nachum, O.; Yang, M.~S.; Lee, L.; Freeman, D.; Guadarrama, S.; Fischer, I.; Xu, W.; Jang, E.; Michalewski, H.; et~al. 2022.
\newblock Multi-game decision transformers.
\newblock \emph{Advances in Neural Information Processing Systems}, 35: 27921--27936.

\bibitem[{Li et~al.(2023)Li, Chong, Stepputtis, Campbell, Hughes, Lewis, and Sycara}]{theory}
Li, H.; Chong, Y.~Q.; Stepputtis, S.; Campbell, J.; Hughes, D.; Lewis, M.; and Sycara, K. 2023.
\newblock Theory of mind for multi-agent collaboration via large language models.
\newblock \emph{arXiv preprint arXiv:2310.10701}.

\bibitem[{Li et~al.(2022)Li, Chen, Jin, Tan, Zha, and Wang}]{pico}
Li, W.; Chen, H.; Jin, B.; Tan, W.; Zha, H.; and Wang, X. 2022.
\newblock Multi-Agent Path Finding with Prioritized Communication Learning.
\newblock arXiv:2202.03634.

\bibitem[{Liang et~al.(2023)Liang, Huang, Xia, Xu, Hausman, Ichter, Florence, and Zeng}]{ecode}
Liang, J.; Huang, W.; Xia, F.; Xu, P.; Hausman, K.; Ichter, B.; Florence, P.; and Zeng, A. 2023.
\newblock Code as policies: Language model programs for embodied control.
\newblock In \emph{2023 IEEE International Conference on Robotics and Automation (ICRA)}, 9493--9500. IEEE.

\bibitem[{Liu et~al.(2023{\natexlab{a}})Liu, Jiang, Zhang, Liu, Zhang, Biswas, and Stone}]{llmp}
Liu, B.; Jiang, Y.; Zhang, X.; Liu, Q.; Zhang, S.; Biswas, J.; and Stone, P. 2023{\natexlab{a}}.
\newblock Llm+ p: Empowering large language models with optimal planning proficiency.
\newblock \emph{arXiv preprint arXiv:2304.11477}.

\bibitem[{Liu et~al.(2023{\natexlab{b}})Liu, Yu, Zhang, Xu, Lei, Lai, Gu, Ding, Men, Yang et~al.}]{agentbench}
Liu, X.; Yu, H.; Zhang, H.; Xu, Y.; Lei, X.; Lai, H.; Gu, Y.; Ding, H.; Men, K.; Yang, K.; et~al. 2023{\natexlab{b}}.
\newblock Agentbench: Evaluating llms as agents.
\newblock \emph{arXiv preprint arXiv:2308.03688}.

\bibitem[{Ma, Luo, and Ma(2021)}]{dhc}
Ma, Z.; Luo, Y.; and Ma, H. 2021.
\newblock Distributed Heuristic Multi-Agent Path Finding with Communication.
\newblock arXiv:2106.11365.

\bibitem[{Ma, Luo, and Pan(2021)}]{dcc}
Ma, Z.; Luo, Y.; and Pan, J. 2021.
\newblock Learning Selective Communication for Multi-Agent Path Finding.

\bibitem[{Okumura(2023)}]{lacam}
Okumura, K. 2023.
\newblock Lacam: Search-based algorithm for quick multi-agent pathfinding.
\newblock In \emph{Proceedings of the AAAI Conference on Artificial Intelligence}, volume~37, 11655--11662.

\bibitem[{Omidshafiei et~al.(2017)Omidshafiei, Pazis, Amato, How, and Vian}]{decmarl}
Omidshafiei, S.; Pazis, J.; Amato, C.; How, J.~P.; and Vian, J. 2017.
\newblock Deep decentralized multi-task multi-agent reinforcement learning under partial observability.
\newblock In \emph{International Conference on Machine Learning}, 2681--2690. PMLR.

\bibitem[{Phan et~al.(2024)Phan, Driscoll, Romberg, and Koenig}]{cactus}
Phan, T.; Driscoll, J.; Romberg, J.; and Koenig, S. 2024.
\newblock Confidence-Based Curriculum Learning for Multi-Agent Path Finding.
\newblock \emph{arXiv preprint arXiv:2401.05860}.

\bibitem[{Radford et~al.(2018)Radford, Narasimhan, Salimans, Sutskever et~al.}]{gpt1}
Radford, A.; Narasimhan, K.; Salimans, T.; Sutskever, I.; et~al. 2018.
\newblock Improving language understanding by generative pre-training.

\bibitem[{Raffel et~al.(2020)Raffel, Shazeer, Roberts, Lee, Narang, Matena, Zhou, Li, and Liu}]{t5}
Raffel, C.; Shazeer, N.; Roberts, A.; Lee, K.; Narang, S.; Matena, M.; Zhou, Y.; Li, W.; and Liu, P.~J. 2020.
\newblock Exploring the limits of transfer learning with a unified text-to-text transformer.
\newblock \emph{Journal of machine learning research}, 21(140): 1--67.

\bibitem[{Rashid et~al.(2020)Rashid, Samvelyan, De~Witt, Farquhar, Foerster, and Whiteson}]{qmix}
Rashid, T.; Samvelyan, M.; De~Witt, C.~S.; Farquhar, G.; Foerster, J.; and Whiteson, S. 2020.
\newblock Monotonic value function factorisation for deep multi-agent reinforcement learning.
\newblock \emph{Journal of Machine Learning Research}, 21(178): 1--51.

\bibitem[{Sartoretti et~al.(2019)Sartoretti, Kerr, Shi, Wagner, Kumar, Koenig, and Choset}]{primal}
Sartoretti, G.; Kerr, J.; Shi, Y.; Wagner, G.; Kumar, T. K.~S.; Koenig, S.; and Choset, H. 2019.
\newblock PRIMAL: Pathfinding via Reinforcement and Imitation Multi-Agent Learning.
\newblock \emph{IEEE Robotics and Automation Letters}, 2378--2385.

\bibitem[{Sharon et~al.(2015)Sharon, Stern, Felner, and Sturtevant}]{cbs}
Sharon, G.; Stern, R.; Felner, A.; and Sturtevant, N.~R. 2015.
\newblock Conflict-based search for optimal multi-agent pathfinding.
\newblock \emph{Artificial Intelligence}, 40--66.

\bibitem[{Skrynnik et~al.()Skrynnik, Andreychuk, Borzilov, Chernyavskiy, Yakovlev, and Panov}]{pogemabench}
Skrynnik, A.; Andreychuk, A.; Borzilov, A.; Chernyavskiy, A.; Yakovlev, K.; and Panov, A. ????
\newblock Pogema: A benchmark platform for cooperative multi-agent navigation, 2024a.
\newblock \emph{URL https://arxiv. org/abs/2407.14931}.

\bibitem[{Skrynnik et~al.(2022)Skrynnik, Andreychuk, Yakovlev, and Panov}]{pogema}
Skrynnik, A.; Andreychuk, A.; Yakovlev, K.; and Panov, A.~I. 2022.
\newblock POGEMA: partially observable grid environment for multiple agents.
\newblock \emph{arXiv preprint arXiv:2206.10944}.

\bibitem[{Sunehag et~al.(2017)Sunehag, Lever, Gruslys, Czarnecki, Zambaldi, Jaderberg, Lanctot, Sonnerat, Leibo, Tuyls et~al.}]{vdn}
Sunehag, P.; Lever, G.; Gruslys, A.; Czarnecki, W.~M.; Zambaldi, V.; Jaderberg, M.; Lanctot, M.; Sonnerat, N.; Leibo, J.~Z.; Tuyls, K.; et~al. 2017.
\newblock Value-decomposition networks for cooperative multi-agent learning.
\newblock \emph{arXiv preprint arXiv:1706.05296}.

\bibitem[{Szot et~al.(2023)Szot, Schwarzer, Agrawal, Mazoure, Metcalf, Talbott, Mackraz, Hjelm, and Toshev}]{etasks}
Szot, A.; Schwarzer, M.; Agrawal, H.; Mazoure, B.; Metcalf, R.; Talbott, W.; Mackraz, N.; Hjelm, R.~D.; and Toshev, A.~T. 2023.
\newblock Large language models as generalizable policies for embodied tasks.
\newblock In \emph{The Twelfth International Conference on Learning Representations}.

\bibitem[{Talebirad and Nadiri(2023)}]{multic}
Talebirad, Y.; and Nadiri, A. 2023.
\newblock Multi-agent collaboration: Harnessing the power of intelligent llm agents.
\newblock \emph{arXiv preprint arXiv:2306.03314}.

\bibitem[{Tan(1993)}]{iql}
Tan, M. 1993.
\newblock Multi-agent reinforcement learning: Independent vs. cooperative agents.
\newblock In \emph{Proceedings of the tenth international conference on machine learning}, 330--337.

\bibitem[{Tobin et~al.(2017)Tobin, Fong, Ray, Schneider, Zaremba, and Abbeel}]{domainrand}
Tobin, J.; Fong, R.; Ray, A.; Schneider, J.; Zaremba, W.; and Abbeel, P. 2017.
\newblock Domain Randomization for Transferring Deep Neural Networks from Simulation to the Real World.
\newblock \emph{CoRR}, abs/1703.06907.

\bibitem[{Touvron et~al.(2023)Touvron, Lavril, Izacard, Martinet, Lachaux, Lacroix, Rozi{\`e}re, Goyal, Hambro, Azhar et~al.}]{llama}
Touvron, H.; Lavril, T.; Izacard, G.; Martinet, X.; Lachaux, M.-A.; Lacroix, T.; Rozi{\`e}re, B.; Goyal, N.; Hambro, E.; Azhar, F.; et~al. 2023.
\newblock Llama: Open and efficient foundation language models.
\newblock \emph{arXiv preprint arXiv:2302.13971}.

\bibitem[{Vaswani et~al.(2017)Vaswani, Shazeer, Parmar, Uszkoreit, Jones, Gomez, Kaiser, and Polosukhin}]{attention}
Vaswani, A.; Shazeer, N.; Parmar, N.; Uszkoreit, J.; Jones, L.; Gomez, A.~N.; Kaiser, {\L}.; and Polosukhin, I. 2017.
\newblock Attention is all you need.
\newblock \emph{Advances in neural information processing systems}, 30.

\bibitem[{Wagner and Choset(2011)}]{Mstar}
Wagner, G.; and Choset, H. 2011.
\newblock M*: A complete multirobot path planning algorithm with performance bounds.
\newblock In \emph{2011 IEEE/RSJ international conference on intelligent robots and systems}, 3260--3267. IEEE.

\bibitem[{Wang et~al.(2020)Wang, Ren, Liu, Yu, and Zhang}]{qplex}
Wang, J.; Ren, Z.; Liu, T.; Yu, Y.; and Zhang, C. 2020.
\newblock Qplex: Duplex dueling multi-agent q-learning.
\newblock \emph{arXiv preprint arXiv:2008.01062}.

\bibitem[{Wang et~al.(2022)Wang, Wei, Schuurmans, Le, Chi, Narang, Chowdhery, and Zhou}]{selfcot}
Wang, X.; Wei, J.; Schuurmans, D.; Le, Q.; Chi, E.; Narang, S.; Chowdhery, A.; and Zhou, D. 2022.
\newblock Self-consistency improves chain of thought reasoning in language models.
\newblock \emph{arXiv preprint arXiv:2203.11171}.

\bibitem[{Wang et~al.(2023)Wang, Xiang, Huang, and Sartoretti}]{scrimp}
Wang, Y.; Xiang, B.; Huang, S.; and Sartoretti, G. 2023.
\newblock SCRIMP: Scalable Communication for Reinforcement- and Imitation-Learning-Based Multi-Agent Pathfinding.
\newblock arXiv:2303.00605.

\bibitem[{Wei et~al.(2022)Wei, Wang, Schuurmans, Bosma, Xia, Chi, Le, Zhou et~al.}]{cot}
Wei, J.; Wang, X.; Schuurmans, D.; Bosma, M.; Xia, F.; Chi, E.; Le, Q.~V.; Zhou, D.; et~al. 2022.
\newblock Chain-of-thought prompting elicits reasoning in large language models.
\newblock \emph{Advances in neural information processing systems}, 35: 24824--24837.

\bibitem[{Wu et~al.(2023)Wu, Tang, Mitchell, and Li}]{smartplay}
Wu, Y.; Tang, X.; Mitchell, T.~M.; and Li, Y. 2023.
\newblock Smartplay: A benchmark for llms as intelligent agents.
\newblock \emph{arXiv preprint arXiv:2310.01557}.

\bibitem[{Yao et~al.(2024)Yao, Yu, Zhao, Shafran, Griffiths, Cao, and Narasimhan}]{tot}
Yao, S.; Yu, D.; Zhao, J.; Shafran, I.; Griffiths, T.; Cao, Y.; and Narasimhan, K. 2024.
\newblock Tree of thoughts: Deliberate problem solving with large language models.
\newblock \emph{Advances in Neural Information Processing Systems}, 36.

\bibitem[{Yao et~al.(2022)Yao, Zhao, Yu, Du, Shafran, Narasimhan, and Cao}]{react}
Yao, S.; Zhao, J.; Yu, D.; Du, N.; Shafran, I.; Narasimhan, K.; and Cao, Y. 2022.
\newblock React: Synergizing reasoning and acting in language models.
\newblock \emph{arXiv preprint arXiv:2210.03629}.

\bibitem[{Yu and Mooney(2022)}]{robotics1}
Yu, A.; and Mooney, R.~J. 2022.
\newblock Using both demonstrations and language instructions to efficiently learn robotic tasks.
\newblock \emph{arXiv preprint arXiv:2210.04476}.

\bibitem[{Zhang et~al.(2023)Zhang, Du, Shan, Zhou, Du, Tenenbaum, Shu, and Gan}]{eagents}
Zhang, H.; Du, W.; Shan, J.; Zhou, Q.; Du, Y.; Tenenbaum, J.~B.; Shu, T.; and Gan, C. 2023.
\newblock Building cooperative embodied agents modularly with large language models.
\newblock \emph{arXiv preprint arXiv:2307.02485}.

\bibitem[{Zheng, Zhang, and Grover(2022)}]{onlinedt}
Zheng, Q.; Zhang, A.; and Grover, A. 2022.
\newblock Online decision transformer.
\newblock In \emph{international conference on machine learning}, 27042--27059. PMLR.

\end{thebibliography}

\end{document}